\documentclass[one-column,showpacs,preprintnumbers,prb,aps]{revtex4}

\usepackage{graphicx}
\usepackage{dcolumn}
\usepackage{amsfonts,amsmath,amssymb,bm}
\usepackage{natbib}
\usepackage[colorlinks]{hyperref}
\setcitestyle{super}

\begin{document}

\title{\bf {Influence of Al-doping on the structural, magnetic, and electrical
properties of La$_{1-x}$Ba$_x$Mn$_{1-x}$Al$_x$O$_3$ ($0\leq x \leq 0.25$)  manganites}}
\date{\today}
\author{J.~Ardashti~ Saleh}
\author{I.~Abdolhosseini~Sarsari$^{*}$}
\author{P.~ Kameli}
\author{H.~Salamati}
\affiliation{a) Department of Physics, Isfahan University of Technology, Isfahan, 84156-83111, Iran\\
}

\newcommand{\etal}{{\em et al}}
\begin{abstract}
We have studied the effect of Al doping on the structural, magnetic
and electrical properties of La$_{1-x}$Ba$_x$Mn$_{1-x}$Al$_x$O$_3$ ($0\leq x \leq 0.25$) 
manganite, annealed in two 750$^oC$ and 1350$^oC$ temperatures.  
The XRD analysis shows that the structures in all samples
have single phase rhombohedral structure with R$\bar{3}$c space group.
The unit cell volume almost decrease with increasing the Al doping in all samples.
The grain growth with increasing annealing temperature  and  Al doping also have been studied.
We observed that, T$_c$  temperature decreases when the Al ion substitute in Mn ion site. 
The magnetic study of the samples via magnetic susceptibility results 
in Griffiths and spin-glass phase for samples doped with aluminium.
Along the resistivity measurement results, the $T_{MIT}$ (metal-insulator) 
transition temperatures decrease and the system
become an insulator. The insulator-metal transition occurs for
L0 sample in near 165K, while this transition is weak for H0 sample due to oxygen
non-stoichiometry.
Using three models viz. 1. Adiabatic small polaron hoping, 2.Variable range hopping, 
and 3. Percolation model, the resistance have been studied.

\end{abstract}

\maketitle
\section{Introduction}

During the last two decades, perovskite manganites are widely studied due
to their interesting physical properties. Perovskite manganite with the
formula A$_{1-x}$B$_x$MnO$_3$ (A is trivalent rare-earth cation such as
La, Sm, Nd, Pr, ..., and B is divalent alkali or alkaline earth
cation such as Sr, Ca, Ba or other vacancies) is any of a variety of
manganese oxides with strongly correlated electrons
\cite{jonker1950,ramakrishnan,lakshmi,phan}.
The doping in A and B sites provide cation size mismatch on ABO$_3$ perovskite manganites,
measured by tolerance factor, that impact on the structures of the polycrystalline samples.
Undoped $LaMnO_3$, as a typical compound, is an antiferromagnetic (AFM)  insulator.
However, when part of La site is substituted
by a divalent metal ion (like Ca, Ba or Sr), the compound  becomes metallic ferromagnetic
(FM) and introduce a ferromagnetic-paramagnetic (PM) transition (Curie temperature).
Also a metal-semiconductor transition near the Curie temperature ($T_C$) could
be found.

The fundamental feature of the magnetic properties and  conductive mechanism  in manganites can
be explained qualitatively, by the double exchange (DE) mechanism, Jahn-Teller distortion and electron-phonon
interactions in manganites. The DE mechanism is responsible for the ferromagnetic state due to transfer of itinerant $e_g$ electron between  the $Mn^{3+}þ-O-Mn^{4+}$þ bond through the $O^{2-}$ ion due to on-site Hund's coupling~\cite{zener1951,anderson,wollan}. Jahn-Teller distortion which  the structure will distort by removing the degeneracy of the $e_g$ orbitals to  stabilizes in the $3d_{3z^2-r^2}$ with respect to the $3d_{x^2-y^2}$ orbitals~\cite{louca1997local}.
There is a direct link
between the Jahn-Teller distortions and the polarons and this coupling is locally present in the
metallic and insulating phases~\cite{louca}.

To understand the  structural, magnetic and transport properties of
La$_{1-x}$A$_x$MnO$_3$ manganites, several studies have been reported on
the doping of A-site with divalent ions~\cite{rodriguez1996cation,
ju2000anomalous,kar2005study,lopez2003intergranular,kameli2008}.
A substitution at Mn (B-site) dramatically affects the structural, magnetic 
and transport properties of manganites. One of these compounds is La$_{1-x}$Ba$_x$MnO$_3$ (LBMO) manganite.
~\cite{narreto2014,aslibeiki2009reentrant,rathod2012structural,abassi2014structural,dhahri2015}.
As trivalent $La^{3+}$ ions are replaced with a divalent $Ba^{2+}$
in $LaMnO_3$ manganite, some of the manganese ion valence  changed
from $Mn^{3+}$ (with the electronic configuration $3d^4$,~$t^3_{2g}$ $e^1_{g}$,~S=2)
to $Mn^{4+}$ (with the electronic configuration $3d^4$,~$t^3_{2g}$ $e^0_{g}$ ,~S=3/2)
due to introducing holes into the this material. These holes permit charge transfer
in the $e_g$ state which is highly hybridized with the oxygen 2p state.
In the following, the manganese ion valence can be change again  by replacing the
trivalent $Al^{3+}$ ions instead of the trivalent $Mn^{3+}$ ion~\cite{krishnan2000}.
This case can cause a large changes in intended manganite.
$Al^{3+}$ does not possess a magnetic moment and  do not participate
in the magnetic interactions. Also the Al ion has a smaller size
($0.535{\AA}$) than that  a Mn ion ($0.645{\AA}$) and makes increase
in the structural stress~\cite{narreto2014,krishnan2000}.
Based on the foregoing, the La$_{1-x}$Ba$_x$Mn$_{1-x}$Al$_x$O$_3$ manganite is one of the most attractive
manganites. The La$_{1-x}$Ba$_x$Mn$_{1-x}$Al$_x$O$_3$ manganite having the $T_C$
and $T_N$ (Neel temperature) in the temperature range of $295-50$ K for doping
range of $0\leq x \leq 0.8$ ~\cite{narreto2014}.

Various models try to explain the  conductive mechanism  in perovskite manganites.
Among these models we point out  the variable range hopping (VRH), the adiabatic
small polaron hoping (ASPH), percolation and Zhang models~\cite{manjunatha2015,zhang}.
The ASPH model offered  for high temperature ($T>\frac{\theta_D}{2}$) resistivity  in the paramagnetic
regime  at temperatures higher than $T_C$. $\theta_D$ is the Debye temperature and will be explained in the following. According  to this model,
the activation energy $E_0$ is the depth of the local
potential barrier of a trapped polaron~\cite{chouket2016}.

The VRH model proposed to explain conduction mechanism for manganites via thermally activated
small polarons in the adiabatic regime ($T<\frac{\theta_D}{2}$). In the VRH model the carriers
are localized by random potential fluctuation.
In percolation model, there is competition between FM and PM phases
~\cite{manjunatha2015}, while according to Zhang model,
a perovskite grain can be divided into a body phase and a surface
phase in which the body phases have higher Curie temperature
and magnetization~\cite{zhang}.

In this work, we investigate the effect of gradual
B-site substitution by the $Al^{3+}$ ion at the Mn-site
in the  La$_{0.80}$Ba$_{0.20}$Mn$_{1-x}$Al$_x$O$_3$
($0\leq x \leq 0.25$) compound.
The changes in the structural, magnetic,
and electrical properties of the
La$^{3+}_{0.80}$Ba$^{2+}_{0.20}$Mn$^{3+}_{0.8-x}$Mn$^{4+}_{0.2}$Al$^{3+}_x$O$^{2}_3$
(x = 0, 0.10, 0.15, 0.20, and 0.25) manganite polycrystalline samples have
been reported by using XRD, ac magnetic susceptibility, and electrical
resistivity measurements.

\section{Experiment details}
The La$_{1-x}$Ba$_x$Mn$_{1-x}$Al$_x$O$_3$ samples (x=0, 0.10, 0.15, 0.20, and 0.25 ) 
have been successfully prepared using sol-gel method.
The required materials with purity of 99.9\% are as follows:

Ba$_2$(NO$_3$)$_2$ (Barium~nitrate), La(NO$_3$).9H$_2$O (Lanthanum~nitrate),
Mn(NO$_3$).4H$_2$O (Manganese~nitrate), Al(NO$_3$)$_3$.9H$_2$O (Aluminum~nitrate),
C$_6$H$_8$O$_7$(Citric~acid), and C$_{10}$H$_{16}$N$_2$O$_8$(Ethylene~diamine~tetra-acetic~acid~or~EDTA).

The nitrates was dissolved in deionized water and mixed with EDTA
and ultimately, PH adjusted to 7.
The chemical equation for x=0, 0.10, 0.15, 0.20, and 0.25 samples in
La$_{1-x}$Ba$_x$Mn$_{1-x}$Al$_x$O$_3$ is as following:\\
\begin{widetext}
[$\alpha$ La(NO$_3$)$_{3} .6$ H$_2$ O$_{(aq)}$+ $\beta$ Ba(NO$_3$)$_2$ $_{(aq)}$+
$\gamma$ Mn(NO$_3$)$_{2}.4$H$_2$O$_{(aq)}$+  $\sigma$ Al(NO$_3$)$_{3}.9$H$_2$O$_{(aq)}$]+
3[$\eta$ C$_6$H$_8$O$_7$.H$_2$O$_{(aq)}$]+ 2[C$_{10}$H$_{16}$N$_2$O$_8$ $_{(aq)}$]
$\rightarrow$ $\phi$ La$_{0.80}$ Ba$_{0.20}$ Mn$_{1-x}$ Al$_{x}$O$_3$ $_{(s)}$+
$\omega$ CO$_{2 (g)}$+$\mu$ H$_2$O$_{(g)}$+ $\theta$ N$_2$ $_{(g)}$
\end{widetext}
where, the Greek letters represent
the amount of each compound.
The molar ratio of acid, EDTA and nitrate materials were taken 1:2:3.
The solution changed to a gel when it exposed to heat for drying the
solution, (the temperature increased to 220$^oC$
with rate of $\frac{1^oC}{6 min}$). 2 hours Milling and 5 hours
sintering at 450$^o$C result in homogenous composition.

Annealing at two different temperatures, 750$^o$C for 5 hours and
1350$^o$C for 48 hours, samples have a different grain size and grain 
boundaries (GBs) in these compounds. LBMO phases are beginning to
grow at temperatures higher than 600$^o$C~\cite{sadighi2013}.
The slab samples were pressed under 354Psi (31.9MPa) pressure with typical
dimensions of about 1.30$\times$3.22$\times$3.15~mm$^3$.

GBs growth, with increasing temperature from 750$^o$C to 1350$^o$C, impact
their magnetic and electrical properties. We demonstrate  the slab samples
with doping level of x=0, 0.10, 0.15, 0.20, and 0.25 that annealed in 750$^o$C labled:
L0, L10, L15, L20, and L25 (L refers to low temperature), while those annealed
in  1350$^o$C labled: H0, H10, H15, H20, and H25 (H refers to high temperature),
respectively.

Ultimately, we studied the structural, magnetic and electrical effects
of  Al doping on Mn sites in
La$_{0.8}$Ba$_{0.2}$Mn$_{1-x}$Al$_x$O$_3$, by using XRD, SEM,
ac susceptibility analysis and measuring the electrical resistance
for two types of composition prepared at different temperatures.
Phase formation and crystal structure of samples were studied by XRD pattern using
Cu-${k\alpha}$ radiation source with wavelength of $\lambda$=1.5406$\AA{}$ in
the 2$\theta$ range from 20$^o$ to 80$^o$, with a step size of 0.05$^o$.

To study  the magnetic properties of the samples, the ac
susceptibility measurements were performed using a Lake Shore
Ac Susceptometer (Model 7000) in 333Hz frequency and 800 A/m field.
Also, we used conventional four probe method for measuring the
electrical resistance by using a Laybold closed cycle refrigerator.

\section{Results and Discussion}

\begin{figure}
\includegraphics*[scale=0.30]{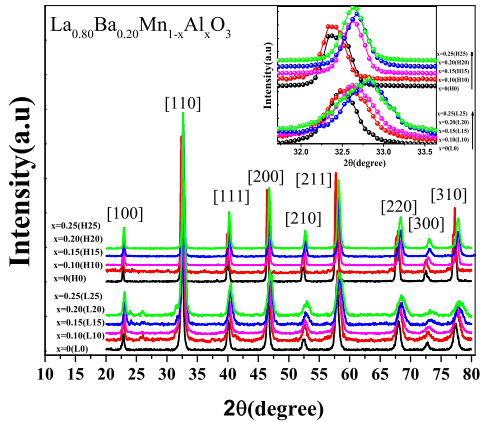}
\caption{\label{xrd}
The XRD pattern of all samples at room temperature that were
sintered at in  750$^o$C and 1350$^o$C. Inset shows zoomed [110] peak.
Can be seen a slightly shift to right hand side in all
peak with increasing Al ion substitution.
}
\end{figure}

\begin{figure}
\includegraphics*[scale=0.30]{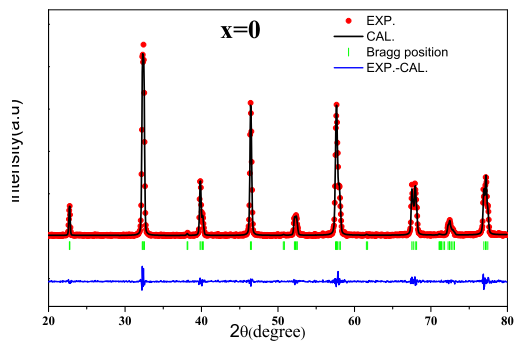}
\caption{\label{reitveld}
The observed and calculated XRD patterns of H0 sample by Reitveld analysis.
Red points (circle) shows the results of through experiments,
black lines obtained from calculation, the blue line shows the
difference between calculation and the results of experience
and eventually green vertical lines, points out the Bragg positions.
}
\end{figure}

Figure~\ref{xrd} shows the  X-ray diffraction (XRD) patterns of 
La$_{0.8}$Ba$_{0.2}$Mn$_{1-x}$Al$_x$O$_3$ (x=0-0.25) at room temperature
that were sintered at 1350$^o$C and 750$^o$C.
The XRD data was analyzed with Rietveld refinement using the FULLPROF
software and Pseudo-Voigt function (for instance H0 sample's rietveld
analysis was shown in Figure~\ref{reitveld}) and  results show that the desired
samples structure has approximately single phase with R$\bar{3}$c
space group belonging to the hexagonal structure where $ a = b\neq c$.
The results obtained  from this analysis were collected in table~\ref{txrd}.

\begin{table*}
\begin{center}
\caption{\label{txrd}
Lattice parameters and volumes, obtained from Rietveld refinement  for all
sample. $R_{wp}$, $R_P$ and $\chi^2$ are the   fitting parameters in Rietveld refinement.
}
\begin{tabular}{|c||c|c|c|c|c|c|c|c|}
\hline
Sample-1350$^o$C     & a=b(\AA{})     &  c(\AA{})     & c/a       & V (\AA{}$^3$)   & $R_{wp}$ &$R_P$  & $\chi^2$  &Space group   \\
\hline
x=0(L0)                  & 5.5240         &13.4690        & 2.4382    & 355.9375        &18.7      &14.9   &1.43       & R$\bar{3}$c  \\
\hline
x=0.10(L10)           & 5.4956         &13.4418        & 2.4459    &351.5755         &24.0      &20.3   &1.48       & R$\bar{3}$c  \\
\hline
x=0.15(L15)           & 5.4931         &13.4995        & 2.4575    &352.7649         &17.6      &13.0   &1.30       & R$\bar{3}$c  \\
\hline
x=0.20(L20)           & 5.4847         &13.4838        &2.4584     &351.2793         &24.2      &17.7   &1.44       & R$\bar{3}$c  \\
\hline
x=0.25(L25)           & 5.4715         &13.4840        &2.4644     &349.5863         &24.9      &18.7   &1.75       & R$\bar{3}$c  \\
\hline
x=0(H0)                 & 5.5456         & 13.4700       &2.4290     &358.7500         &18.8      &11.6   &1.01       & R$\bar{3}$c  \\
\hline
x=0.10(H10)          & 5.5372         &13.4720        &2.4329     &357.7200         &29.7      &23.2   &1.21       & R$\bar{3}$c  \\
\hline
x=0.15(H15)          & 5.5192         &13.4350        & 2.4342    & 354.4400        & 16.9     &10.1   &1.10       & R$\bar{3}$c  \\
\hline
x=0.20(H20)          & 5.5019         &13.4130        & 2.4378    & 351.6400        & 17.9     &12.7   &1.11       & R$\bar{3}$c  \\
\hline
x=0.25(H25)          & 5.4539         &13.4590        & 2.4499    & 351.7500        & 18.2     &11.7   &1.18       & R$\bar{3}$c  \\
\hline
\end{tabular}
\end{center}
\end{table*}

\begin{figure*}
\includegraphics[scale=.40]{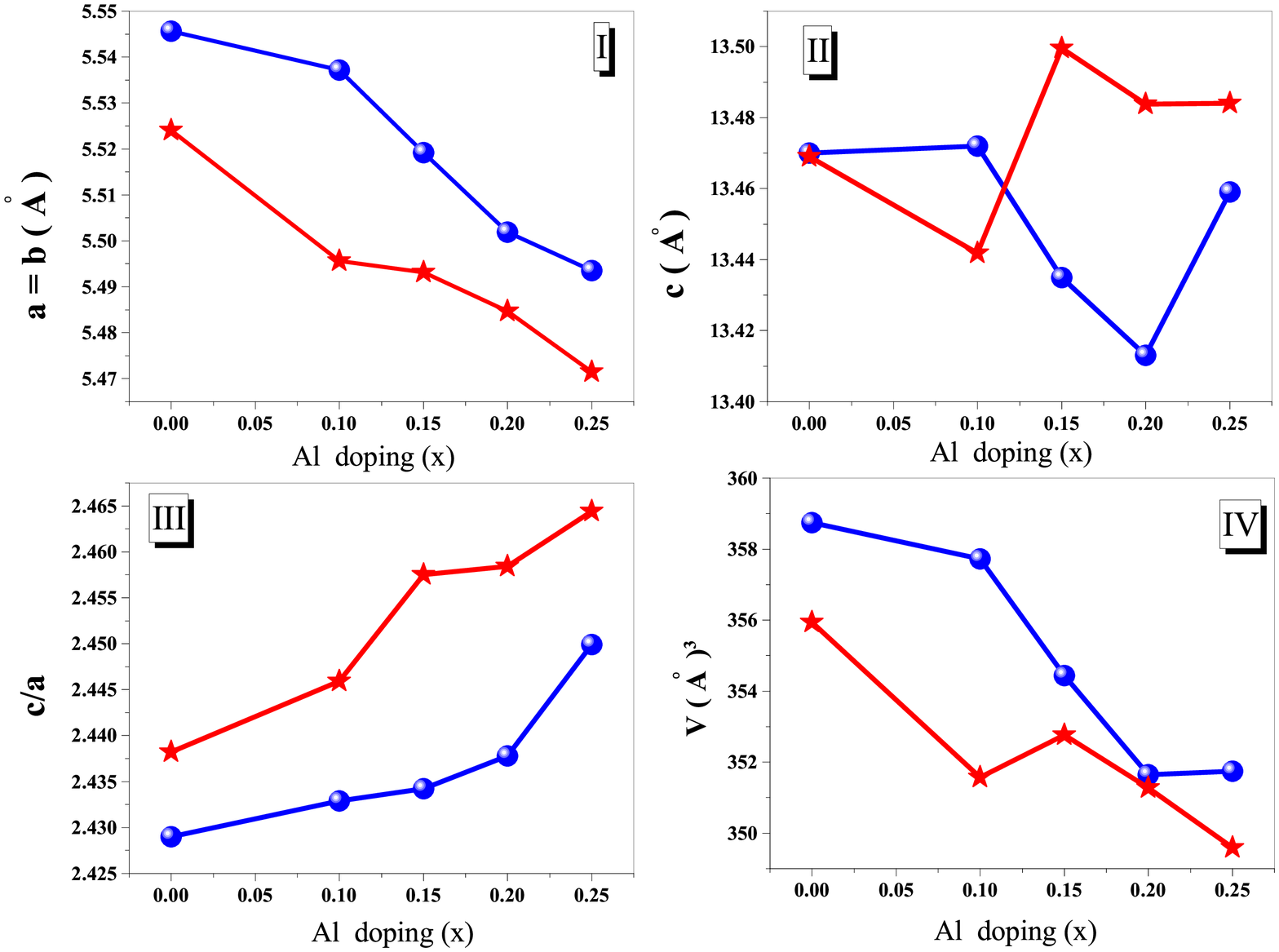}
\caption{\label{parameter}
Variation of (I) lattice parameters  a and b, (II) lattice parameters c,
(III) c/a ratio and (IV) unit cell volume,as a function of Al doping in
La$_{0.8}$Ba$_{0.2}$Mn$_{1-x}$Al$_x$O$_3$ (x=0-0.25). The red star and blue circle
points refer to samples that annealed in 750$^o$C and 1350$^o$C, respectively.
}
\end{figure*}

Figure~\ref{parameter} shows the changes in lattice parameters
with increasing aluminum doping and  annealing temperature.
The $a$ and $b$ parameters decrease, $c$ parameter, doesn't show  regular behavior, $c/a$
parameter increases  and the unit cell volume decreases
(except L15 and H25) with  increasing Al doping.
The  $c/a$ ratio represents a further increase in the length
direction c than a direction (c and a are the unit cell parameters).
It seems that reducing the unit cell volume is due to the fact that Al ions are smaller than Mn ions.
(The radii of Al ion is 0.535~\AA{} and the  Mn ion is 0.645~\AA{})~\cite{narreto2014,krishnan2000}.
By comparing the samples annealed at higher temperatures, it
can be observed that all of lattice parameters  increases
with increasing annealing temperature. The  reason for increasing
in lattice parameters  is  the increasing concentration ratio
of $ Mn^{3+}/Mn^{4+}$. The Goldschmidt radius of $Mn^{3+}$
(0.070nm) is much greater than that of $Mn^{4+}$(0.052 nm)~\cite{zhang2}.
So this increase in the size of the lattice parameters can be
attributed to  the increase of $Mn^{3+}$  ions due to increased
oxygen deficiency with increasing annealing temperature~\cite{zhang2}.

\begin{figure*}
\includegraphics[scale=.50]{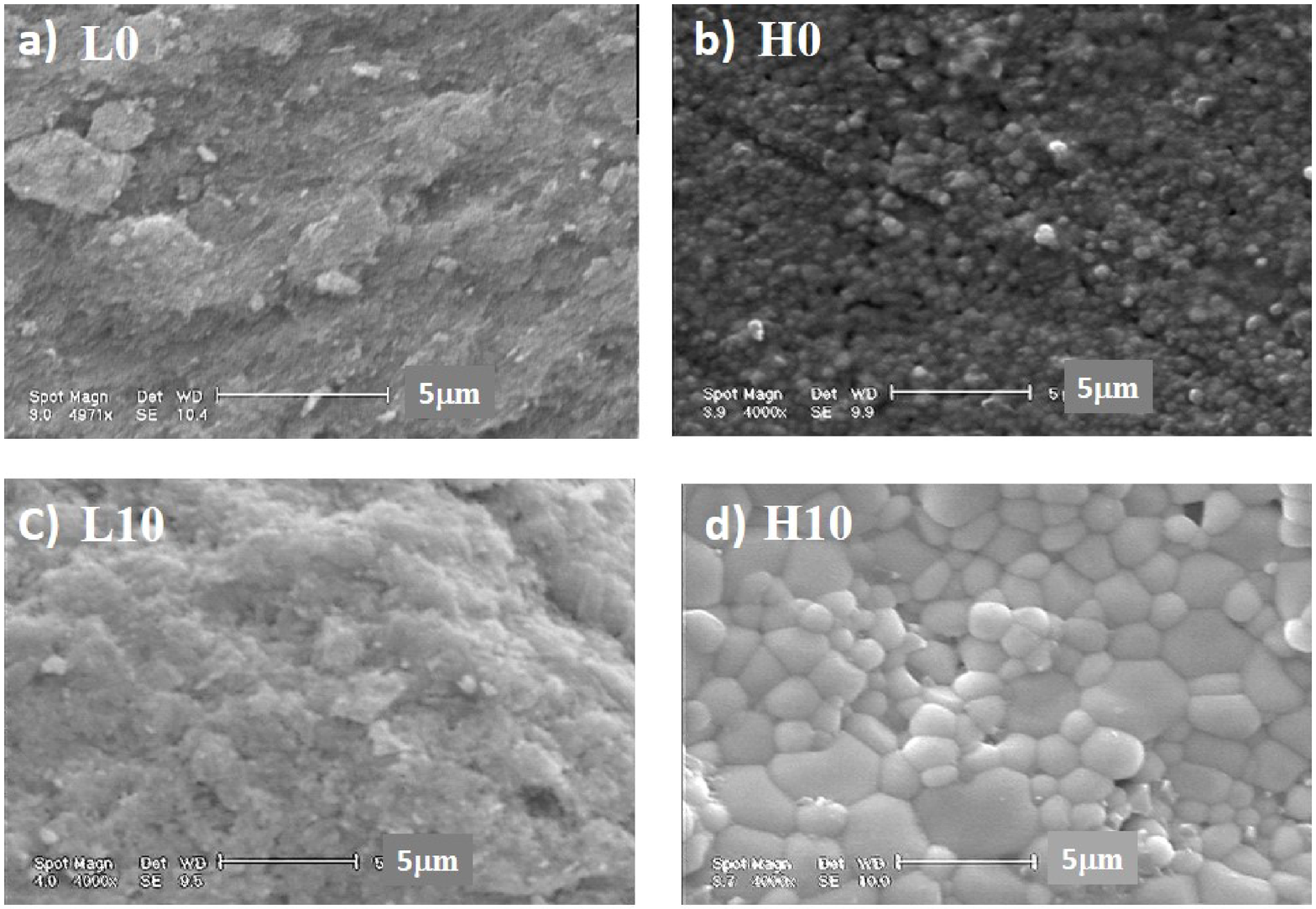}
\caption{\label{sem1}
SEM micrograph of x=0 and x=0.10 samples, a) L0, b) H0, c)
L10 and d) H10 samples (La$_{0.8}$Ba$_{0.2}$Mn$_{1-x}$Al$_x$O$_3$)
revealing the surface morphology and the particle size distribution
at different sintering temperatures and Al doping.
}
\end{figure*}

The SEM images of  slab samples for x=0 and x=0.10 (L0, H0, L10,
and H10 samples) are shown in figure~\ref{sem1}. Comparing the
L0 to H0  and L10 to H10 in figure~\ref{sem1}, the effect of
annealing temperature on these samples can be well understood.
The grains in samples L10 and H10 has grown more than samples L0
and L10, respectively. The average grain sizes of these samples
are smaller than 5 $\mu m$.

The samples annealed in 1350$^o$C (H0 and H10) are more
homogeneous than samples that annealed  in  750$^o$C (L0 and L10).
In general, decrease of grain surface energy by increasing the
annealing temperature, is known as a driving force for
migration of grain growth~\cite{khzouz2011,yue2010}.

Grain size and their boundaries  have impact on the magnetic and
electrical properties of manganites~\cite{kameli2008}.
 It is observable that the average
particle size of the samples increases by increasing the Al doping.
It seems that doping  of Al, cause the grain boundaries have more
driving force  for grain growth.

\subsection{Magnetic properties}
Measuring ac magnetic susceptibility, the magnetic properties of all samples studied.
La$_{0.80}$Ba$_{0.20}$MnO$_3$ manganites show different magnetic
transitions such as PM-FM or PM-AFM
transitions and also FM to spin-glass state transition at different temperatures.
~\cite{sadighi2013,narreto2014,trukhanov2003,chukalkin2005}.
The heat treatment, level of  synthesized temperature,
annealing atmosphere and ..., could affect
susceptibility of the samples, for a certain doping
~\cite{sadighi2013,jiang2010,chu2010}.

\begin{figure}
\includegraphics*[scale=0.30]{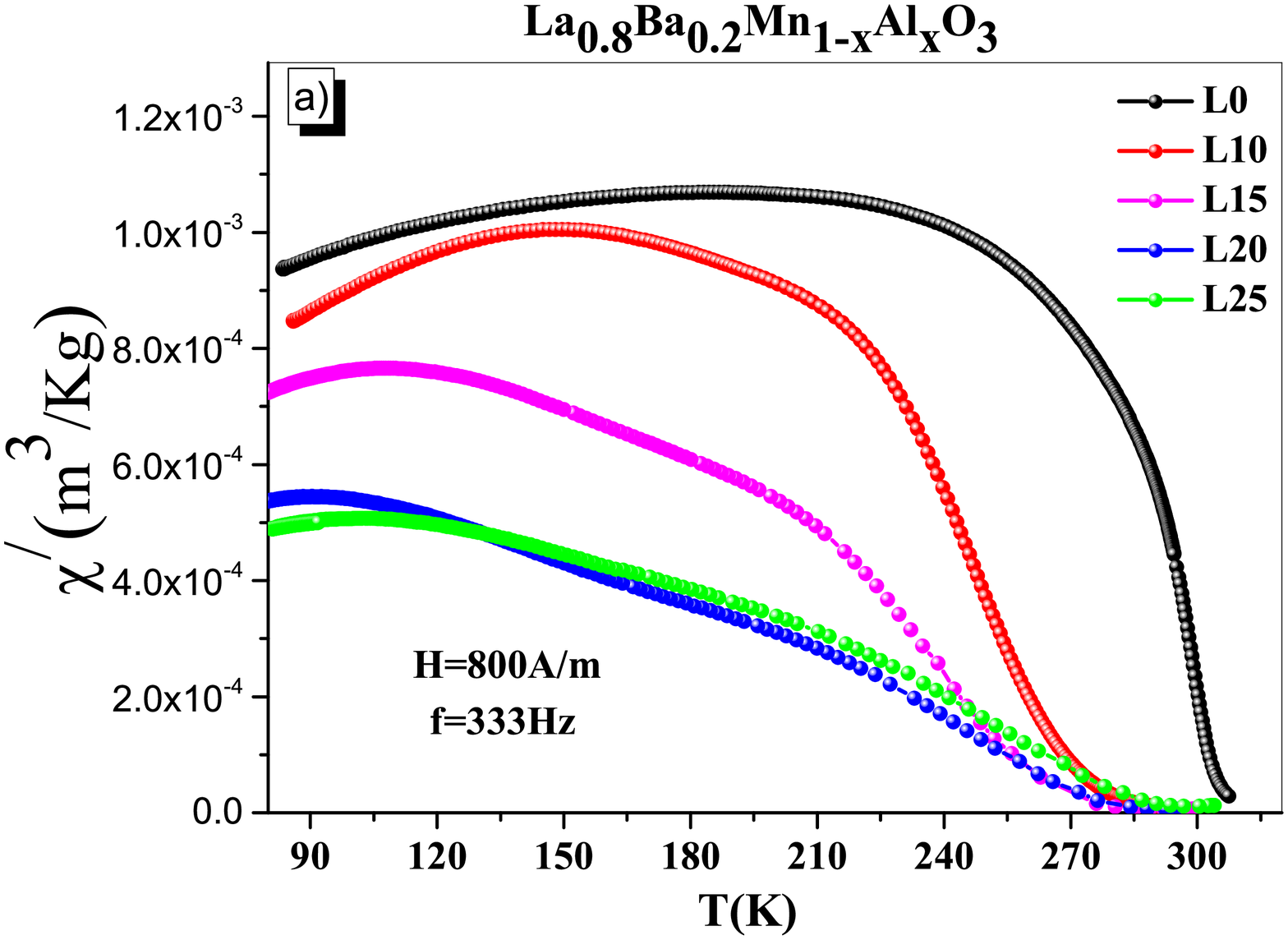}
\includegraphics*[scale=0.30]{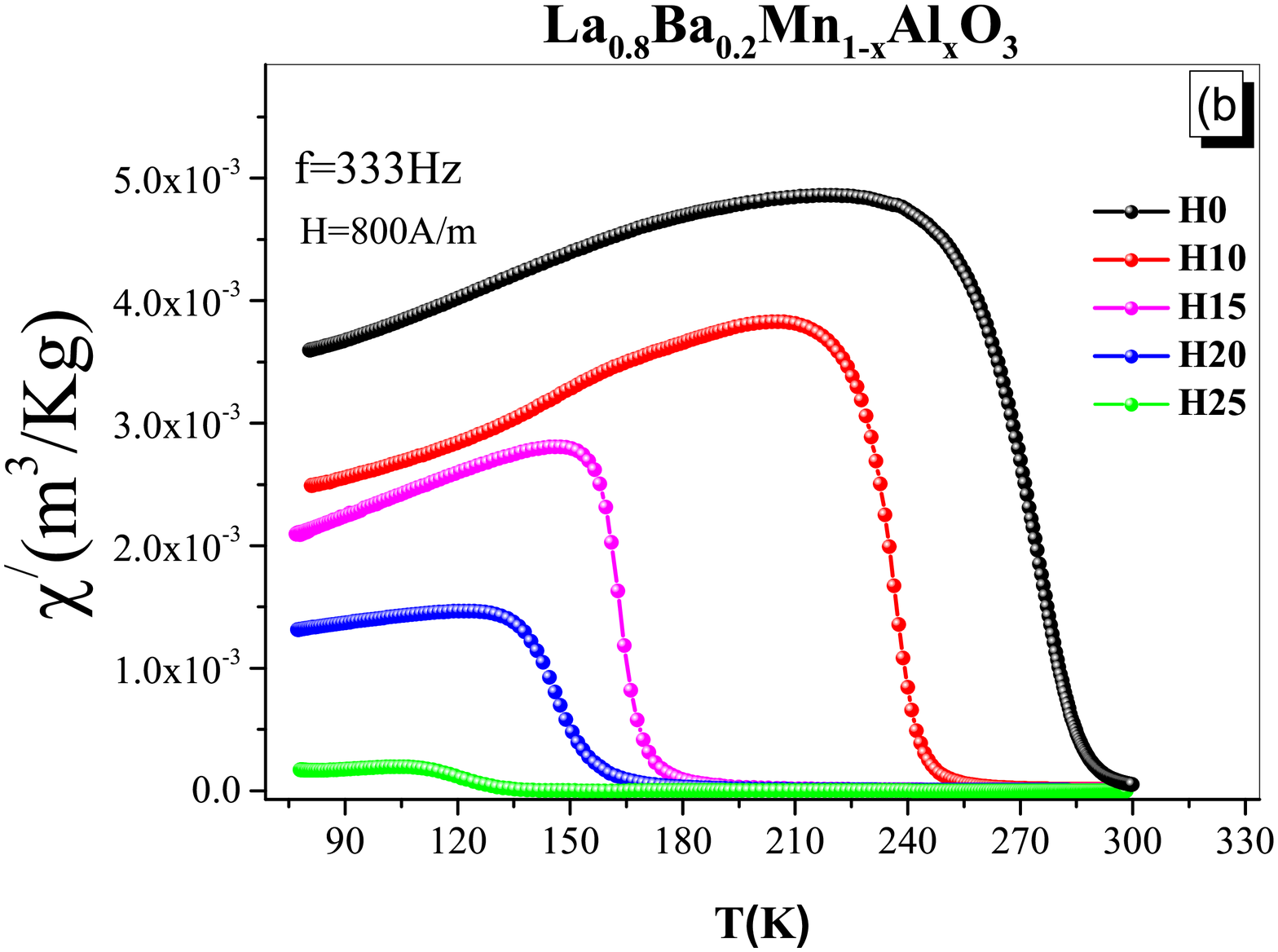}
\caption{\label{real-ac-sus}
Temperature dependence of (a) samples  that annealed in 750$^o$C and (b)
samples  that annealed in 1350$^o$C for x = 0-0.25 samples in a
magnetic field of 10 Oe and frequency of 333 Hz.
}
\end{figure}

Figure~\ref{real-ac-sus} shows  the real parts of ac susceptibility for all samples.
With decreasing temperature, magnetic entropy overcome
to the thermal entropy and therefore below  a certain temperature (Curie temperature, $T_c$)
 the samples show FM behavior.
It can be seen that the Curie temperature decreases with increasing doping concentration.
The decrease of the $T_c$ with Al substitution has many reason such as 1- local
cutoff magnetic interaction between the spins of $t_{2g}$ electrons~\cite{qin2003}
.2- $Al^{3+}$ does not posses a magnetic moment and thus it should not participate
in the magnetic interaction. This cause the ferromagnetic area reduction in $ac$
susceptibility Vs temperature figure. 3- $Al$ ion has smaller size than $Mn$ ion
 and will cause an increase in structural stress with influence the magnetic and
electronic properties of the mixed valence manganites.

\begin{figure}
\includegraphics*[scale=0.35]{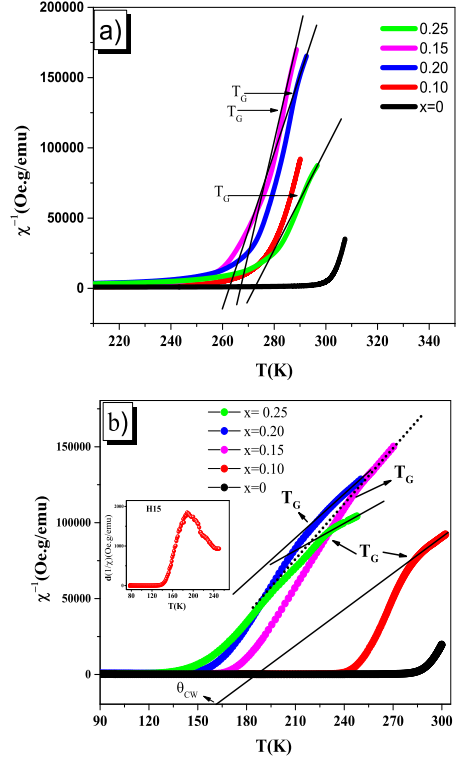}
\caption{\label{grifiskoli}
Temperature dependence of the inverse magnetic susceptibility
for a series of La$_{0.8}$Ba$_{0.2}$Mn$_{1-x}$Al$_x$O$_3$($0\leq x \leq 0.25$)
at H=10 Oe. a)and b) shows the inverse ac susceptibility obtained
from magnetization measurements at H=10Oe for samples that annealed
in 750$^o$C and 1350$^o$C,respectively.  All samples shows the deviation
from the Curie Weiss law (exept L0, L10, and H0). Griffith temperature (T$_G$)
is shown by arrows. $\theta_{CW}$ is the Curie Weiss temperature that shown
for H10  as an example.Notice that the Griffiths temperature $T_G$ is defined
as the maximum of the $(d(1/\chi)/dT)(T)$ curves that shown as an example
for H15 sample  in the inset of part b.
}
\end{figure}

The $\chi^{'}$ shows downward slop below $T_c$  for all samples
( see Fig \ref{real-ac-sus}).
Decreasing the temperature cause the immediate appearance of domain structure
(The domain structure of ferromagnets  is a result of minimizing the free energy)
below T$_c$. However, by more reduction in temperature, samples lose their
response to the ac magnetic field which could be responsible for the diminution of $\chi^{'}$
(real part of the ac susceptibility ) below T$_c$~\cite{levin2001}.
This $\chi^{'}$ behavior can be justified by the change of domain wall movement
(by applying small fields such as 10 Oe) and domain magnetization
rotation (by applying large fields) in an ac applied magnetic field~\cite{levin2001}.
Samples H10, H15, H20, H25 (samples annealed in 1350$^o$C)  and
samples L15, L20 and L25 (samples annealed in 750$^o$C) shows  deviation
from the Curie Weiss law (see Fig.~\ref{grifiskoli}). The Curie Weiss law
is $\chi=\frac{C}{T-\theta_{cw}}$, where C is Curie constant and $\theta_{cw}$ is
Curie-Weiss temperature. The $\theta_{cw}$ represents the molecular interaction
between the moments.

This diversion proposed the short-range FM interactions or cluster
spins formation~\cite{narreto2014}. In general, deviation of $\chi^{-1}$,
represents the Griffiths phase (GP) due to presence of FM clusters.
Increasing temperature, these clusters participate in the paramagnetic phase,
while above a certain temperature, that called Griffiths temperature,
they totally disappear. The GP means the simultaneous
presence of short-range FM clusters in the
PM region, that is due to the dispersion of FM
spin clusters in the PM domain~\cite{bray1987}.
The Griffiths temperature ($T_G$) range is $T_c\leq T \leq T_G$~\cite{tong2008griffiths}.
We used the Eq~\ref{Tg}, to fit the intended experimental data. Also, the percentage of
GP are obtained from Eq~\ref{darsad}~\cite{salamon2002}.

\begin{equation}\label{Tg}
\chi^{-1}\propto (T-T^{rand}_c)^{1-\lambda}
\end{equation}

Where, $T^{rand}_c$ is the critical temperature of the random
ferromagnet where susceptibility diverges ($T^{rand}_c > T_c$), and $\lambda$ is
a positive quantity between 0 and 1~\cite{phong2015}.
Zero refers  to the PM regime.

\begin{equation}\label{darsad}
 GP\%=\frac{T_G-T_C}{T_C}\times 100\%
 \end{equation}

We have obtained the $T^{rand}_c$ and $\lambda$ parameters from equation~\ref{Tg}
and the percentage of GP,  from equation~\ref{darsad} 
listed in table~\ref{grifis}.

In many reports, the  Griffiths phase was observed in Mn-site doping of
$LaMnO_3$ perovskites~\cite{abassi,joshi2009existence,phong2015}.
Disorder caused by A~\cite{rama2004site} and B site doping~\cite{pramanik}
and also controlling and elimination  of oxygen vacancy~\cite{zhang2},
enhance the Griffiths state in a system.
The decreasing in $\lambda$ with $Al^{3+}$ content indicates the
further decrease in the GP properties and  reveals that the size
of the magnetic cluster decreases. It can be observed an increase
in the quantity of $\lambda$  for samples  annealed at lower
temperatures and irregular behavior for samples annealed at
high temperature.

\begin{figure}
\includegraphics*[scale=0.35]{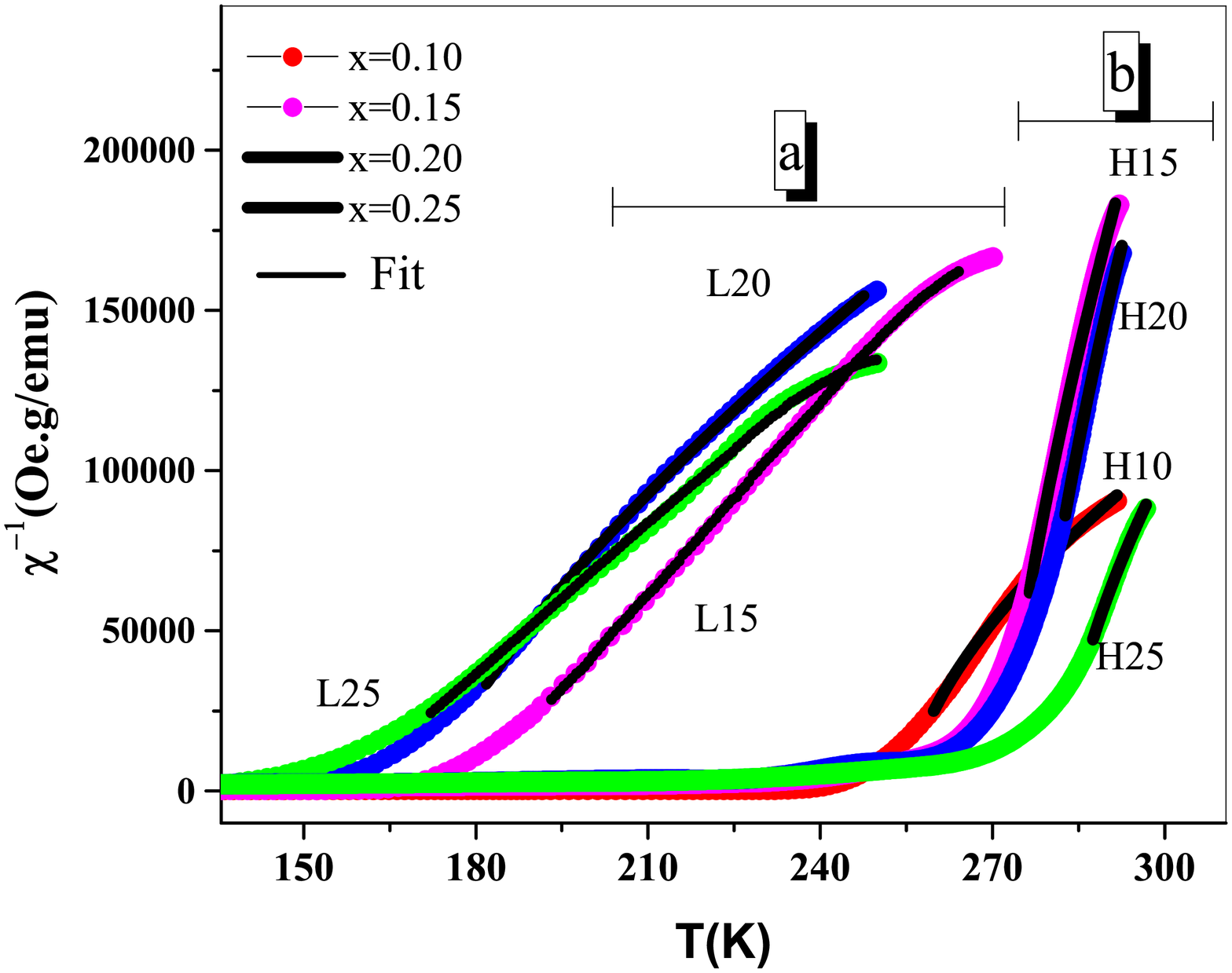}
\caption{\label{fitkoligrifith}
}Inverse of  magnetic susceptibility versus temperature for all
samples of La$_{0.8}$Ba$_{0.2}$Mn$_{1-x}$Al$_x$O$_3$($0\leq x \leq 0.25$)
at H=10 Oe. a) and b)  shows the samples that annealed in 1350$^o$C and
750$^o$C,respectively. The solid lines in the (a) and (b) part, show
the fit with the power law equation $\chi^{-1}\propto (T-T_G)^{1-\lambda}$.
\end{figure}

\begin{table*}
\begin{center}
\caption{\label{grifis}
Parameters derived from fitting experimental curve for All
sample(except L0,L10 and H0) from Eq~\ref{rho1} and Eq~\ref{rho2}.
$T_c$ is Curie temperature, $\theta_p$ refers to Weiss temperature,
$T^{rand}_c$  refers to the disorder dependent FM ordering temperature,
$T_G$ refers to the Griffiths temperature,$\lambda$ refers to the
Griffiths exponent and $GP \times 100\%$ refers to the temperature
range of Griffith phase. $ADj.R^2$ demonstrates the accuracy of
matching equations~\ref{Tg} and ~\ref{darsad} with the experimental data.
}
\begin{tabular}{|c||c|c|c|c|c|c|c|}
\hline
Sample-1350$^o$C    & T$_c$(K)    &$T^{rand}_c$ &$\theta_{CW}$(K)& T$_g$(K)  & $\lambda$  & $GP\cdot \%$ &  $Adj\cdot R^2\%$ \\
\hline
x=0(L0)             & 302         &....         &298             & ....      & ....       &....          &....               \\
\hline
x=0.10(L10)         & 246         &....         &274             & ....      &....        &....          &....               \\
\hline
x=0.15 (L15)        & 242         &272          &264             & 288       &0.286       &19.0          &99.78              \\
\hline
x=0.20(L20)         & 245         &277          &258             &285        &0.305       &16.3          &99.85              \\
\hline
x=0.25(L25)         & 272         &282          &266             & 290       & 0.377      &6.6           &99.84              \\
\hline
\hline
x=0(H0)             & 274         &....         &277             & ....      & ....       &....          &....               \\
\hline
x=0.10(H10)         & 236         & 256         &161             & 275       &0.389       &16.5          &99.50              \\
\hline
x=0.15 (H15)        & 162         &184          &130             & 254       &0.130       &56.7          &99.68              \\
\hline
x=0.20(H20)         & 146         &172          &99              &209        &0.264       &43.1          &99.94              \\
\hline
x=0.25 (H25)        & 119         &161          &14              &233        &0.139       &95.7          &99.57              \\
\hline
\end{tabular}
\end{center}
\end{table*}

\begin{figure}
\includegraphics*[scale=0.35]{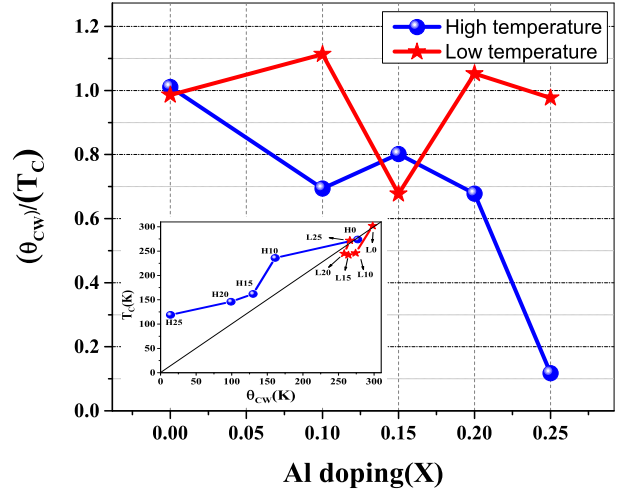}
\caption{\label{tetapi}
 This graph shows the  $\frac{\theta_{CW}}{T_c}$  vs Al doping for all sample.
 The incet shows the  $T_c$ vs $\theta_{CW}$.It can be seen that the difference
 between $T_c$ and $\theta_{CW}$   is very less for samples that annealed in 750$^o$C.}
\end{figure}

Figure~\ref{tetapi} shows the  $\frac{\theta_{CW}}{T_c}$  vs Al doping
for all samples. The inset of this figure shows the $T_c$ vs $\theta_{CW}$.
 According to these results (Table~\ref{grifis} and figure~\ref{tetapi}),
we can point out that; with $Al^{3+}$  doping, amount  of $\theta_{CW}$
reduced that would means weakening of the DE interaction.
~\cite{phong2015}.
It is seen that  the difference between $T_c$ and $\theta_{CW}$
(the values of $\frac{\theta_{CW}}{T_c}$)  increases with increasing
Al dopping for samples annealed at 1350$^o$C
due to the  destruction of the FM order by the AFM
interactions of the next-nearest neighbors~\cite{arbuzova2008magnetic}.
In other words, the relationship between $T_c$ and $\theta_{CW}$ when $\frac{\theta_{CW}}{T_c}<1$
indicate the dominance of the AFM
interactions in the  La$_{0.8}$Ba$_{0.2}$Mn$_{1-x}$Al$_x$O$_3$ (H0, H10, H15, H20, and H25) samples.
This case occurs in reverse order in  samples that  annealed at 750$^o$C (L0, L10, L15, L20, and L25 samples).

Also we have observed spin glass (SG) behavior in sample H15
(sample with doping x=0.15 and annealed in 1350$^o$C). Note that we believe that this
effect could be seen in samples with higher doping, however just been brought
for H15 sample as an example.
One of the features of SG system is the dependence of its ac susceptibility
on the applied field and frequency. In this sample, this behavior results in
a sharp drop in the real part of the susceptibility at low temperatures,
and the appearance of a peak in its imaginary part.
The frequency- independent peaks termed as Hopkinson peaks are typical feature
in many FM materials. But the second peak position shifts to higher
temperatures with increasing frequency. So, to verify SG state presence,
we have measured ac susceptibility of  La$_{0.8}$Ba$_{0.2}$Mn$_{1-x}$Al$_x$O$_3$
(x=0.15) sample on the constant field and different frequencies.

\begin{figure}
\includegraphics*[scale=0.65]{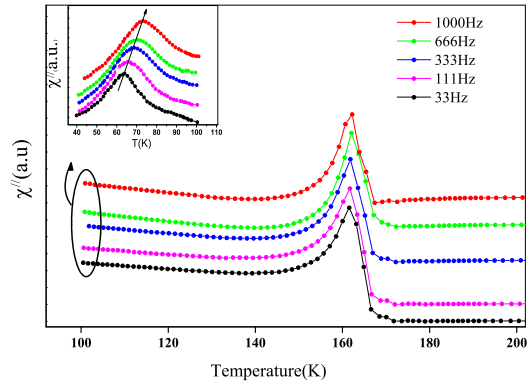}
\caption{\label{width-ac-sus}
The imaginary part of the susceptibility for H15 sample that measured at different frequencies.
The inset shows the evolution of the peak by the increase of the frequency.
}
\end{figure}

Figure~\ref{width-ac-sus} shows the temperature dependence of the imaginary part
(Out of phase) of ac susceptibility at different frequencies.
The temperature dependences of susceptibility of this sample measured
in an applied field of 10 Oe, after cooling in the zero field (ZFC).
As can be seen from Figure~\ref{width-ac-sus}, there is SG behavior
below a certain temperature (see inset in Figure~\ref{width-ac-sus}).

The first peak temperature that is frequency independent, represents the Curie
temperature, while the second peak temperature, represents the freezing temperature
$T_f$, that shifts towards higher temperatures with increasing frequency.
This change by frequency is taken as the smoking gun for spin glasses.
For further investigation of the SG nature at x=0.15 doping, we have checked
the $T_f$'s dependence on frequency by conventional critical slowing down
model which is as follows:

\begin{equation}\label{spin}
 f=f_0(\frac{T_f-T_g}{T_g})^{z\nu}
\end{equation}

\begin{figure}
\includegraphics*[scale=0.55]{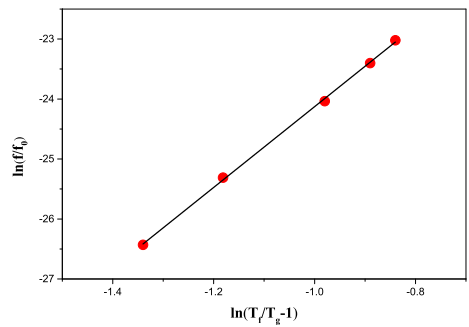}
\caption{\label{fit spin}
Ln-Ln plot of the reduced temperature ($T_f$ /$T_g$ - 1) versus frequency for H15 sample.
}
\end{figure}

Where, $T_g$ is the dc value of $T_f$ for $f\rightarrow0$, $f_0$ is a constant
in order of $10^9$-$10^{13}$ and $z\nu$ is dynamic critical exponent.
The Fig~\ref{fit spin} shows best fit of this model. The estimated values
of $z\nu=6.7$ and $T_g =54K$ are within the realm of three dimensional SG system.

\subsection{Electrical  properties}

Electrical transport measurement is one of the best methods for investigating the electrical
properties of materials such as manganites. Thus to check the electrical resistance
of samples, we used the closed circuit refrigerator and four probe method.

\begin{figure}
\includegraphics*[scale=0.35]{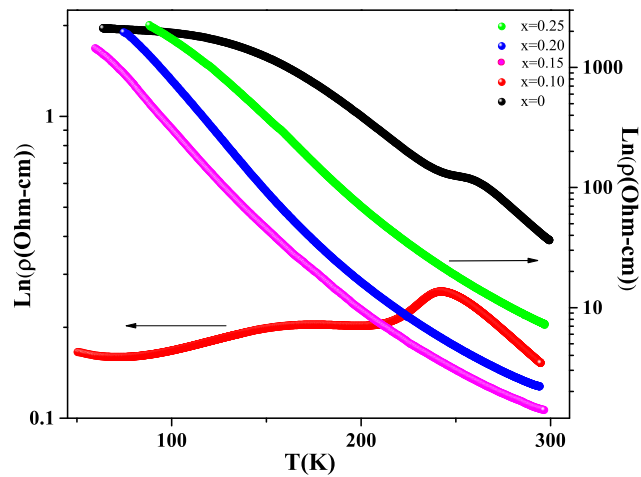}
\caption{\label{rho1350}
Dependence of resistivity on temperature for
La$_{0.8}$Ba$_{0.2}$Mn$_{1-x}$Al$_x$O$_3$ samples that annealed in high temperature.
Resistivity increases with increasing aluminum doping among the high temperature samples.
Only H0 and H10 samples show metal-insulator transition.}
\end{figure}

\begin{figure}
\includegraphics*[scale=0.35]{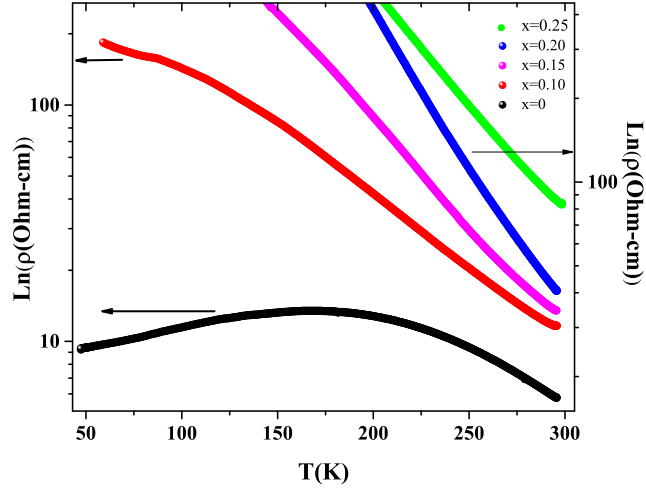}
\caption{\label{rho750}
Dependence of resistivity on temperature for
La$_{0.8}$Ba$_{0.2}$Mn$_{1-x}$Al$_x$O$_3$ samples that annealed in low temperature.
Resistivity increases with increasing aluminum doping among the  low temperature samples.
Only L0 sample show metal-insulator transition.
}
\end{figure}

Figure~\ref{rho1350} and~\ref{rho750} show the electrical resistivity  of all samples.
According to figure~\ref{rho1350}, for x=0 concentration, a poor metal-insulator transition
can be seen at temperatures of about 259K that is corresponding to FM
transition. This behavior is in agreement with La$_{0.81}$Ba$_{0.19}$MnO$_3$ in ref~\cite{ju2000anomalous}.
This is probably due to that sample x=0,  is  approximately on the boundary of the
insulator and metal state~\cite{ju2000anomalous,narreto2014}.
The further doping shows insulator behavior (x=0.19), due to the oxygen vacancy or losing oxygen
(According to figure~\ref{rho750}, the ferromagnetic transition to
the insulating state is completely occurred, and this transition becomes weak with
increasing annealing temperature). This transition is in agreement with work Nagabhushana
et al~\cite{nagabhushana2007}.

In reference~\cite{abdelmoula2000}, the effect of  oxygen  non-stoichiometry has been discussed
and it  has been shown that by increasing the amount of $\delta$ in $O_{3-\delta}$,
the La$_{0.7}$Ba$_{0.3}$MnO$_{3-\delta}$ manganite shows insulation behavior.
But  figure~\ref{rho750}, shows metal-insulators transition for x=0 in 170K.
The reason for this issue can be attributed to the oxygen deficiency
~\cite{abdelmoula2000,lopez2003,nagabhushana2007}.
By increasing the amount of oxygen deficiency, the ratio 
of Mn$^{3+}$/Mn$^{4+}$ is reduced, which leads to decrease the 
transmission $e$$_g$ interaction electrons.
Studies have shown that the Curie temperature, T$_c$, depends on the mobility
of e$_g$ electrons, and the ratio of Mn$^{3+}$/Mn$^{4+}$,  that is a key component
to understanding the magnetoresistance effect and paramagnetic metal-ferromagnetic
semiconductor transition~\cite{abdelmoula2000,jin1994}.
According to the Zener double exchange model, the transfer of itinerant e$_g$
electron between Mn$^{3+}$ and Mn$^{4+}$ ions through the O$^{2-}$ ion,
causing ferromagnetic interaction due to on-site Hund's coupling~\cite{anderson1955}.
When there are  oxygen deficiency, the Mn$^{3+}$/Mn$^{4+}$ percent are changed.
Also, By increasing the content of oxygen deficiency, the e$_g$ electrons hopping
probability and ferromagnetic region are reduced, that leads to decrease in the
Curie temperature~\cite{yang2004}.

In other justification, annealing at higher temperature, the samples H0 lose more
oxygen than L0 sample. This case is clear  in Figure~\ref{rho750} in which 
there is a metal-isulator transition in 168K.
With increasing annealing temperature, grain size increased and the surface to volume
ratio decreased. At the same deoxygenated material,  grain boundaries
which effect by level of aluminium substitution on manganese site,
play crucial role. According to Zhang model, the grain's surface is chaos and the
curie temperature  and metal-insulator transition  temperature are less than the core phase~\cite{zhang}.
Thus, the sample L0 has more surface area than H0, and as a result, the total
Curie temperature is decreased because the Curie temperature of the surface
phase are less.

The electrical resistance is insulating for samples x=0.15, 0.20, and 0.25 (H15, H20, and H25).
Increasing the non-magnetic $Al^{3+}$ ion instead of the magnetic $Mn^{3+}$ ions in
lanthanum based perovskite manganites, reduces metal properties. The
aluminium doping, as an impurity, decreases
the polaron conductivity and increases the insulating properties of the system.
Also increasing  Al substitution (with smaller ionic radii  size ) instead of the  Mn ion,
led to increase stress. Such a structural changes in the lattice strain and deformations,
affect the $Mn^{3+}$-O$Mn^{4+}$ bond angle and length~\cite{narreto2014}.
In addition, with increasing further doping, the  tunnelling process at the grain
boundaries (that acts as a potential barrier for charge carriers), are weakening.
Also by doping the $Al^{3+}$, the content of $Mn^{3+}$  are reduced progressively
and thus power of DE reduced and this makes the ferromagnetic
state weaker and system become insulator.

\begin{figure}
\includegraphics*[scale=0.35]{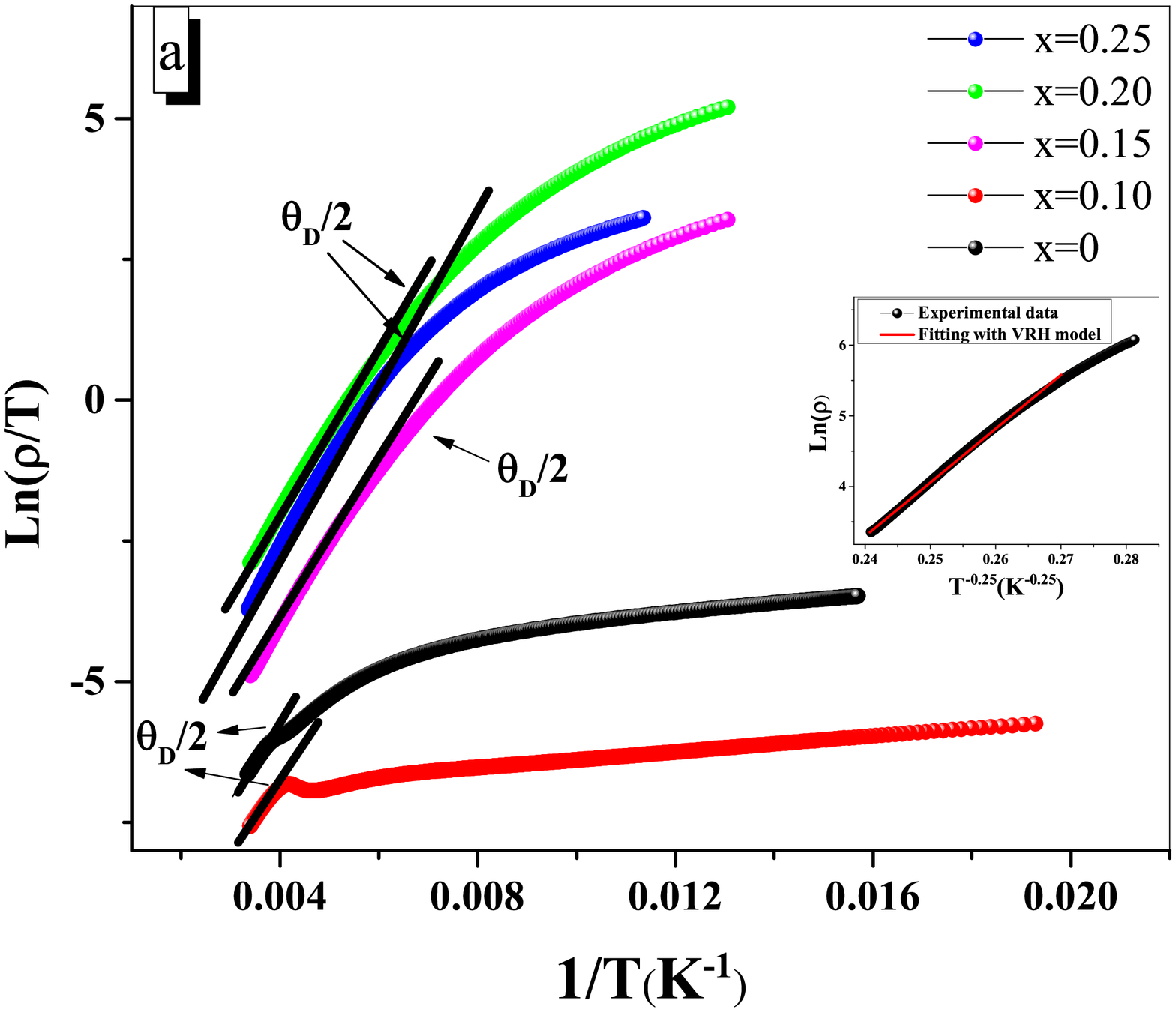}
\includegraphics*[scale=0.35]{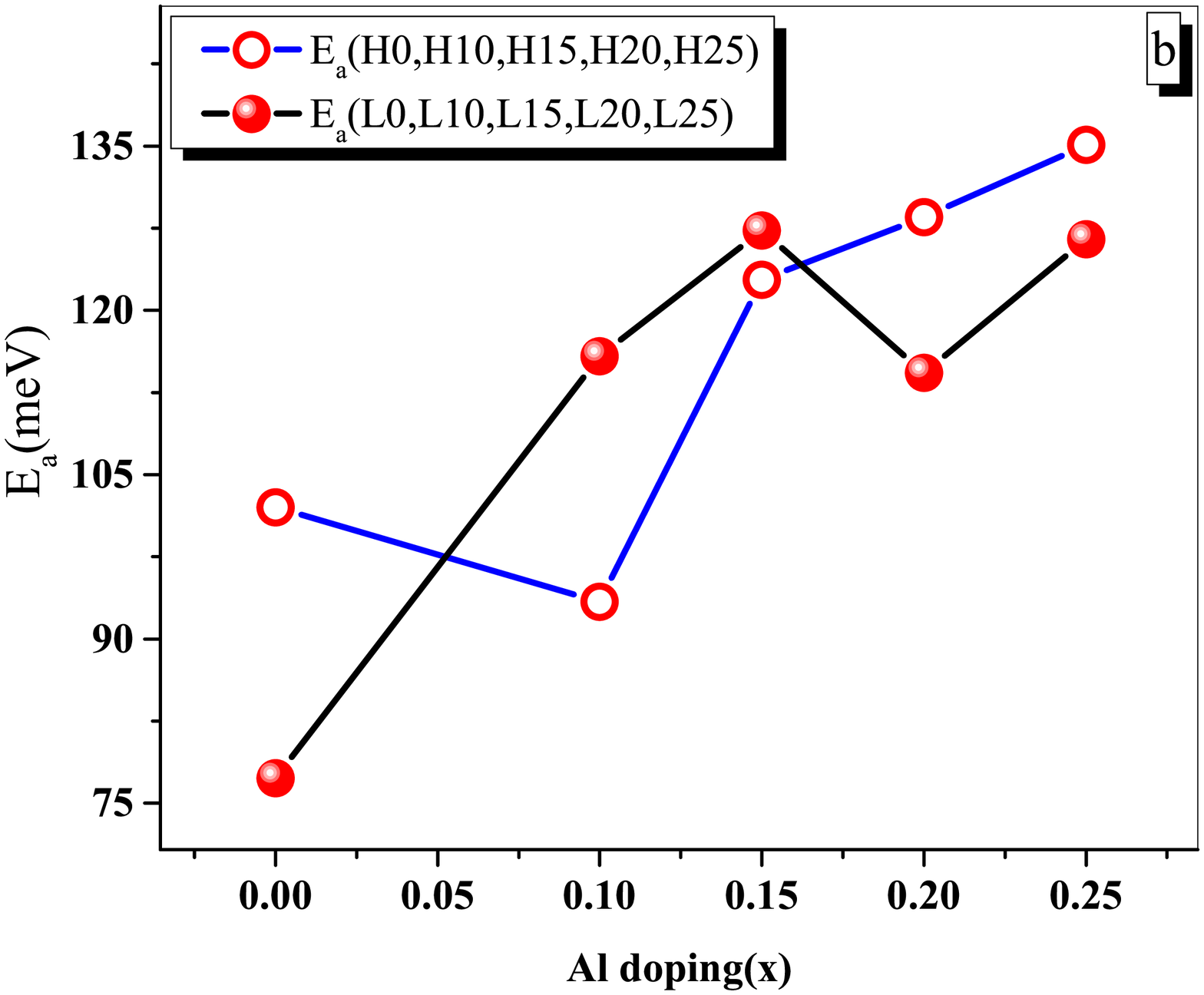}
\caption{\label{ea}
 (a) Theoretical fit of high temperature resistivity data for
 dependence of the  ln$\rho/T$  versus 1/T. The solid lines represent
 the best fit of linear function. The inset of figure ~\ref{ea}$a$ shows 
 the VRH model for H20 sample. (b)Activation energy $E_a$ as a
function of x for  La$_{0.8}$Ba$_{0.2}$Mn$_{1-x}$Al$_x$O$_3$. 
}
\end{figure}

In a further analysis of electrical resistance, we used three
ASPH, VRH, and percolation models.
We used ASPH model to check the
desired electrical properties of samples
in zero applied magnetic field (see Figure~\ref{ea}: It is shown only for high annealed samples).
In this model, which is also known as the Emin-Holstein theory of adiabatic
small polaron hopping model ~\cite{emin}, the conductivity data are
dominated by the thermally activated hopping of small polarons in the high
temperature ($T>T_{MIT}$) insulating  phase~\cite{mott1979}:

\begin{equation}\label{rho1}
 \rho=\rho_{\alpha}T exp(\frac{E_a}{K_BT})
 \end{equation}

In Eq~\ref{rho1}, parameter $\rho_{\alpha}$ is the coefficient independent of T.
$E_a$  and $K_B$ are the activation energy and Boltzmann constant, respectively.
Also, $\rho_{\alpha}=\frac{2K_B}{3ne^2a^2\upsilon_{ph}}$ is the residual resistivity, where
$e$ is the electronic charge, $n$ is the density of charge carriers, $a$ is the site-to-site
hopping distance, and $\upsilon_{ph}$ (in the order of $10^{13}Hz$ ) is the longitudinal optical
phonon frequency~\cite{dhahri2015}.
Also, we used $h\upsilon_{ph}=K_B\theta_{D}$ to calculate the
optical phonon frequency. Here  $\theta_{D}$ is the Debye temperature
that was estimated from  the experimental data in ln$\rho/T$  versus 1/T
plots by choosing the deviation point from linearity behavior.
We calculated $E_a$ for all  samples  in above $T_{MIT}$ as a function
of x using the equation~\ref{rho1}, and plotted the results in Figure~\ref{ea}.
Figure~\ref{ea}$a$ clearly shows a linear dependence of
ln($\rho/T$) on 1/T above $T_{MIT}$ for high temperature annealed samples.
It can be seen  from Figure~\ref{ea}$b$, that $E_a$ has a minimum value of
about 93.4 meV for H10 and 71meV for L0 samples. $E_a$ increases with increasing
$x$ for high temperature annealed samples  and increases with  increasing
x in the $0<x<0.15$ range and then decrease for L20 and L25 samples.

\begin{figure}
\includegraphics*[scale=0.35]{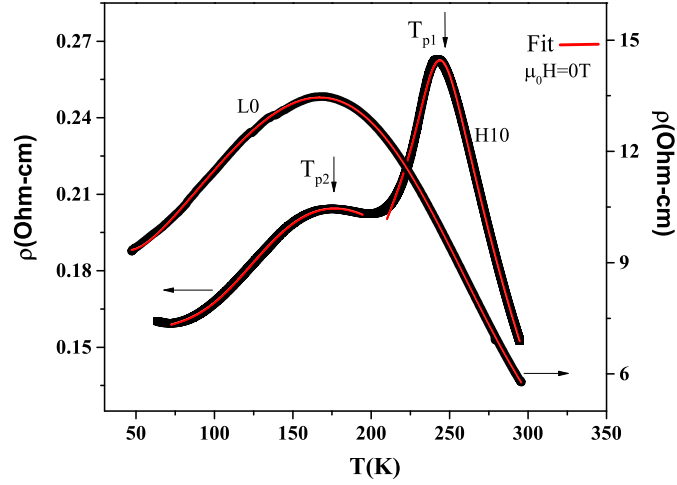}
\caption{\label{fit0.10}
Temperature dependence of the resistivity of
La$_{0.8}$Ba$_{0.2}$Mn$_{1-x}$Al$_x$O$_3$ for L0 and H10
samples. The red solid line
corresponds to the data fitted by Eq.~\ref{rhokoli}.}
\end{figure}

Also to understand the temperature dependence of electrical resistivity
data in semiconducting region, we used $VRH$ model.
The expression for $VRH$  model can  be written as:

\begin{equation}\label{vrH}
 \rho=\rho_{0}T exp(\frac{T_{0}}{T})^{1/4}
 \end{equation}

where $\rho_{0}$ is the  Mott residual resistivity and $T_{0}$ is the
Mott temperature which is expressed as $T_{0}=18\alpha^3/K_BN(E_F)$.
Here $N(E_F)$ is the density of states (DOS) near the Fermi level,
$\alpha$  is localization length and $K_B$ is the Boltzmann constant.
Inset of Fig~\ref{ea}$a$ shows the fitting of the experimental data of
ln$\rho$ versus $T^{1/4}$ for H25 sample as an example. The values for
$Adj\cdot R^2\%$ in Eqs.~\ref{rho1} and~\ref{vrH} are  very close to 1.
The table~\ref{vrh} includes the parameters of the $VRH$ model.

\begin{table*}
\begin{center}
\caption{\label{vrh}
Fitting parameters obtained from resistivity data for
La$_{0.8}$Ba$_{0.2}$Mn$_{1-x}$Al$_x$O$_3$  samples by
using VRH model. $\theta_{D}$ is the Debye temperature, $\upsilon_{ph}$
is the longitudinal optical phonon frequency, $T_{0}$ is the
Mott temperature and  $N(E_F)$ is the density of states (DOS) near the Fermi level.}
\begin{tabular}{|c|c|c|c|c|c|||c|c|c|c|c|}
\hline
Sample             & (L0)      & (L10)      & (L15)         & (L20)     &(L25)     &(H0)         &(H10)       &(H15)      &(H20)      &(H25)      \\
\hline
\hline
$E_a(meV)$         & 77.3        & 115.8      & 127.3            &114.3      &126.5         & 102         &93.4        &122.8       &128.5      &135.1  \\
\hline
$\theta_{D}(K)$     & 535.5      &  467.3     & 437.6           & 475.3      &476.3           &526.3       &514.1       &293.8      &335.6      &363.6   \\
\hline
$\upsilon_{ph}\times10^{13}$      & 1.115      & 0.937           &0.911           &0.989           &0.973     & 1.09       & 1.07           &0.611            &0.698           &0.751\\
\hline
$T_{0}\times10^{7}$       &...           & 2.6          &3.3           &1.6       &0.973     &...          &....      &4.4       &3.2    &1.6   \\
\hline
$N(E_F)\times10^{24}$      &...           & 0.666          &0.507           &1.03       &0.67     &...          &....      &0.386      &0.5311    &1.06   \\
\hline
\end{tabular}
\end{center}

\begin{center}
\caption{\label{rhotable}
Matching parameters of electrical resistance graph
of samples L0(x=0) and H10(x=0.10), by equation~\ref{rhokoli}. The $T_{p1}$ and $T_{p2}$ are shown in Fig.~\ref{fit0.10}.}
\begin{tabular}{|c|c|c|c|c|c|c|c|c|}
\hline
Sample code                          & $\rho_0$         &$\rho_2\times10^{-4}$       &$\rho_{4.5}\times 10^{-12}$     &$\rho_\alpha$      &$E_a/K_B $ &$T_c^{mod}(K)$ & $Adj\cdot R^2\%$\\
&$(\Omega cm)$      &$(\Omega cmK^{-2})$          &$(\Omega cmK^{-4.5})$      &$(\Omega cmK^{-4})$   &$(meV)$ &   &  \\
\hline
$LBMA_{0}$-L0                     &15.15        & 22.6           &7.28           &0.07        &660.48  &168   & 99.95   \\

\hline
$LBMA_{0.10}-H10-T_{p1}$             &0.95        & 1.45           &3.53           &3.19        &796.36 &242   & 99.31 \\

\hline
$LBMA_{0.10}$-H10-$T_{p2}$            &0.15       & 0.14           &14.95           &6.14$\times10^{-4}$        &301.52 &172  & 99.69  \\

\hline
\end{tabular}
\end{center}
\end{table*}

Also to elucidate the transport properties in the whole temperature range, especially
around the transition peak, we used percolation model for sample L0 and H10
(Because only these two dopings  shows  $T_{MIT}$ in between all samples). In this model, the total resistivity is a sum of contributions
from PM and FM regions, and  at any temperature, $ \rho$, is determined
by the change of the volume fractions of both regions.
In other words, in the percolation model, it is assumed that the
ferromagnetic metallic and paramagnetic insulating regions are electrically linked in series and there
exists a competition between these phases around the metal
insulator transition temperature. The resistivity for the entire temperature range
expressed as~\cite{li2002competition,phong2009electrical}:

\begin{equation}
 \rho=f\rho_{FM}+(1-f)\rho_{PM}
 \end{equation}

Here $\rho_{fm}$ is the temperature dependent electrical
resistivity data that is given by:
\begin{equation}\label{rho2}
\rho_{FM}= \rho_{0}+\rho_2T^2+\rho_{4.5}T^{4.5}
\end{equation}

Where the temperature independent part $ \rho_0$ is the resistivity due to domain/grain boundary
and refers to defects scattering. The $\rho_2T^2 $ term represents the electrical resistivity due to the
electron-electron scattering process. The term $\rho_{4.5}$ is a combination of
electron- electron, electron-phonon and electron-magnon scattering processes
~\cite{schiffer1995,snyder1996}.
Also, $\rho_{PM}$ is given by Eq.~\ref{rho1}.
As well as, $f$ is the volume fraction of the FM phase and $(1-f)$ is the volume
fraction of the PM phase~\cite{li2002}. Volume fractions of the FM and PM phases satisfy the Boltzmann distribution:

\begin{equation}
f=\frac{1}{1+exp(\frac{\Delta U}{K_BT})}
\end{equation}

Where $\Delta U$ is the energy difference between FM and PM states and may be expressed as
$\Delta U\approx-U_{0}(1-\frac{T}{T_{c}^{mod}})$.
In this expression, parameters U$_0$ is an   energy difference of the quasi particles  in the  phase separated FM
and PM states for a temperature well below $T_{c}^{mod}$   temperature.
$T_{c}^{mod}$ means a temperature in the vicinity of which the resistivity has a maximum value.
(It is near or equal to $T_c$) Thus the percolation model's formula given as:

\begin{widetext}
\begin{equation}\label{rhokoli}
 \rho=(\frac{1}{1+exp(\frac{-U_{0}(1-\frac{T}{T_{c}^{mod}})}{K_BT})})(\rho_0+\rho_2T^2+\rho_{4.5}T^{4.5})+
 (\frac{exp(\frac{-U_{0}(1-\frac{T}{T_{c}^{mod}})}{K_BT})}{1+exp(\frac{-U_{0}(1-\frac{T}{T_{c}^{mod}})}{K_BT})})
 (\rho_{\alpha}T exp(\frac{E_P}{K_BT}))
\end{equation}
\end{widetext}

From equation~\ref{rhokoli}, one can find that:\\
\checkmark $f=0$ for $T\gg T^{mod}_{c}$, $f=1$ for $T\ll T^{mod}_{c}$ and $f=f_c=1/2$ at $T=T_{c}^{mod}$.\\
\checkmark $1-f=1$ for $T\gg T^{mod}_{c}$ ,$1-f=0$ for $T\ll T^{mod}_{c}$ and $1-f=f_c=1/2$ at $T=T_{c}^{mod}$.\\
where $f_c$ is the percolation threshold. As a consequence, when $f<f_c$ , the sample remains semiconducting and when
$f>f_c$  it becomes metallic.
Parameters $\rho_0$, $\rho_2$, $\rho_{4.5}$, $\rho_{\alpha}$ and $E_p$ have been introduced previously.
The  red  color solid line in
 Figure~\ref{fit0.10}  shows the fitting results for the $\rho$-$T$ curves obtained at
zero field for the samples L0 and H10.
All the available parameters obtained for samples that shows an insulator - metal transition (L0, H10), listed in 
Table~\ref{rhotable}, by using Eq~\ref{rhokoli}  to fit the experimental data.
These Values are consistent with values  of previous works~\cite{dhahri2015,li2002competition,manjunatha2015}.
It is worth mentioning that the percolation model is suitable to explain
simultaneously the electrical transport of La$_{0.8}$Ba$_{0.2}$Mn$_{1-x}$Al$_x$O$_3$
(for samples that shows metal-insulator transition) in both
ferromagnetic and paramagnetic areas.

\section{CONCLUSION}
In summary, the effect of the Al substitution on the structural, magnetic, and electrical properties of
La$_{1-x}$Ba$_x$Mn$_{1-x}$Al$_x$O$_3$ ($0\leq x \leq 0.25$) manganites were investigated
by XRD, Ac susceptibility and electrical resistivity measurements.
We prepared two samples for each concentration, at  temperatures of 750$^o$C and 1350$^o$C
via sol-gel method . The XRD rietveld refinement indicates rhombohedral single
phase structure with R$\bar{3}$c space group.
The lattice parameters and volume decrease  with  Al doping.
The larger average grain size obtained due to aluminium doping and increasing annealing
temperature. The ac magnetic susceptibility measurements show that
the transition from paramagnetic (PM) to ferromagnetic (FM) phase at the Curie
temperature, $T_C$, decreases from 274 K down to 119 K with increase in the Al
doping level from x = 0 to x = 0.25 for samples  annealed at 1350$^o$C.
In addition, the spin glass state exists in the x = 0.15 and higher level doping
sample. This behavior
indicates that the substitution of Al weakens the double exchange (DE) process.
Also we observed a grifiths  phase for some samples that annealed at 750$^o$C
and 1350$^o$C.
The temperature dependence of resistivity, $\rho(T)$, indicates that
by increasing the Al doping level for samples  annealed at 1350$^o$C,
the metal-insulator transition  observed only for x=0 and x=0.10,
while for x=0 there is a very weak metal-insulator transition, eventually
the heavily doped samples become insulators.
But for x=0 sample that annealed
at 750$^o$C, we observed the  metal-insulator transition  near 180K. We have attributed this effect  to oxygen deficiency.
To further analysis the dependence of resistivity on temperature we used
three models (adiabatic small polaron hopping, variable range hopping
and percolation model) for some  samples.
The $\rho(T)$ curve for the samples  that shows  metal-insulator
transition  was fitted with the  percolation model, while the insulating
region fitted with the adiabatic small polaron hopping at
Paramagnetic  state at $T>\frac{\theta_D}{2}$ ($\theta_D$, Debye temperature) and
the variable range hopping  models at $T<\frac{\theta_D}{2}$.

\section{ACKNOWLEDGEMENTS}
This work was supported by Isfahan University of Technology (IUT).
\bibliography{Ref}
\end{document}